%% file: 2ndQNote5.tex
\documentclass[11pt]{article}
\usepackage{typearea}\typearea{12}
\usepackage{amsmath,amsfonts,amssymb,mathtools}
\usepackage[dvipdfmx]{graphicx}
\input{macros}

\newcommand{\phr}{\ph(\bsr)}
\newcommand{\Rt}{\bbR^3}
\newcommand{\idr}{\int d^3\bsr}
\newcommand{\idrN}{\int d^3\bsr_1\cdots d^3\bsr_N}
\newcommand{\roN}{\bsr_1,\ldots,\bsr_N}

\newcommand{\hpsi}{\hat{\psi}}
\newcommand{\hpsid}{\hat{\psi}^\dagger}

\newcommand{\hak}{\hat{\mathfrak{a}}}
\newcommand{\hakd}{\hat{\mathfrak{a}}^\dagger}
\newcommand{\hcB}{\hat{\calB}}

\renewcommand{\tCd}{\hat{a}^\dagger}
\renewcommand{\tC}{\hat{a}}
\renewcommand{\bss}{\boldsymbol{s}}
\newcommand{\bskappa}{\boldsymbol{\kappa}}

\renewcommand{\hilb}{\calH_1}
\renewcommand{\calHwf}{\calH}
\renewcommand{\cong}{=}

\begin{document}
\noindent
{\bf\Large Introduction to the ``second quantization'' formalism for non-relativistic quantum mechanics
 \\
 \large
 A possible substitution for Sections 6.7 and 6.8 of Feynman's ``Statistical Mechanics"}
\par\bigskip

\renewcommand{\thefootnote}{\fnsymbol{footnote}}
\noindent
Hal Tasaki\footnote{
Department of Physics, Gakushuin University, Mejiro, Toshima-ku, 
Tokyo 171-8588, Japan.\\
%\href{mailto:hal.tasaki@gakushuin.ac.jp}{hal.tasaki@gakushuin.ac.jp}
}
\renewcommand{\thefootnote}{\arabic{footnote}}\setcounter{footnote}{0}

\begin{quotation}
\small\noindent
This is a self-contained and hopefully readable account of the method of creation and annihilation operators (also known as the Fock space representation or the ``second quantization'' formalism) for non-relativistic quantum mechanics of many particles.
%\footnote{%
%Japanese translation is available: {https://www.gakushuin.ac.jp/~881791/pdf/2ndQNoteJ.pdf}}
Assuming knowledge only of conventional quantum mechanics in the wave function formalism, we define the creation and annihilation operators, discuss their properties, and introduce corresponding representations of states and operators of many-particle systems.\footnote{%
{\em Note to experts:}\/
In particular, we here {\em derive}\/ the (anti)commutation relations of the creation and annihilation operators rather than simply declaring them.
In this sense, our approach is quite close to that of Feynman.
But we here focus on the action of creation/annihilation operators on general $N$ body wave functions, while Feynman heavily uses Slater-determinant-type states from the beginning.
We hope that our presentation provides a better perspective on this formalism.
}
As the title of the note suggests, we cover most topics treated in sections~6.7 and 6.8 of Feynman's\footnote{%
A friend of mine pointed me out that this happens to be the centennial year of Richard Feynman's birth.
Let me declare that this small article is to celebrate his 100th birthday!
} textbook\footnote{%
I wrote this note for undergraduate students in our group studying Feynman's textbook.
The idea was that they could skip these two somewhat complicated sections by studying this note (and they indeed did so).
} \cite{Feynman}.
As a preliminary, we also carefully discuss the symmetry of wave functions describing indistinguishable particles.

We note that all the contents of the present note are standard, and the definitions and derivations presented here have been known to many.
Although the style of the present note may be slightly more mathematical than standard physics literature, we do not try to achieve full mathematical rigor.\footnote{%
The mathematically minded reader might be bothered by our heuristic treatment of the operators $\hpsi(\bsx)$ and $\hak(\bsk)$.
Our treatment can be made rigorous by using suitable advanced concepts. 
}
%\vspace{0.15cm}
%\noindent
%{\em Note to experts}\/: I like Feynman's approach to the ``second quantization'' formalism, where he does not simply declare the (anti)commutation relations of the creation/annihilation operators, but tries to derive them through explicit construction of these operators that starts from conventional quantum mechanics.
%Nevertheless I was never fully satisfied with his logic, and have always felt that it can be improved.
%In particular his heavy use of Slater determinant type states may give a wrong impression that the creation/annihilation operators are 
%
%
\end{quotation}

\tableofcontents

%%%%%%%%%%%%%%%%%%%%%%%%%%%%%%%%%%%%%%
\section{Wave functions of many particles}
\label{ss:WF}
As a preliminary, we discuss the description of many-particle states in quantum systems in terms of wave functions.
Although most of the material in this section is standard and covered in elementary quantum mechanics courses, we present a detailed and careful discussion of the symmetry of wave functions for indistinguishable particles.

\paragraph*{Single particle}
We start by recalling the standard quantum mechanical description of a single particle, such as an electron or an atomic nucleus.
A state (at an instantaneous moment) of a particle in the three-dimensional space is described by a wave function $\phr$, which is a complex-valued function of the position
%\footnote{%
%For simplicity we neglect the spin of the particle.
%} 
$\bsr=(x,y,z)\in\Rt$.
The wave function $\phr$ satisfies the condition of square integrability:
\eq
\idr\,|\phr|^2<\infty
\lb{si}
\en
We denote by $\hilb$ the set of all wave functions satisfying \rlb{si}, and call it the single-particle Hilbert space.
Recall that the function which is identically 0 is contained in $\hilb$, but does not describe any physical state.

Let us use ``mathematicians' notation'' and denote the whole wave function $\phr$ (where $\bsr$ runs over the whole space $\Rt$) as $\bsph$.
(We shall reserve the bra-ket notation for the Fock space description.)
For two wave functions $\bsph,\bspsi\in\hilb$, we define their inner product by
\eq
\sbkt{\bsph,\bspsi}\coloneqq{}\idr\,\{\phr\}^*\,\psi(\bsr).
\en
The norm of a state $\bsph\in\hilb$ is defined as $\snorm{\bsph}\coloneqq{}\sqrt{\sbkt{\bsph,\bsph}}$.

\paragraph*{Distinguishable particles}
Before dealing with indistinguishable particles, we consider a  system of mutually distinguishable particles.

First consider a system of two particles, which we call particle 1 and particle 2.
Suppose that the particle 1 is in a single-particle state $\bsph\in\hilb$ and the particle 2 in $\bspsi\in\hilb$.
By denoting the positions of the particles 1 and 2 as $\bsr_1$ and $\bsr_2$, respectively, the whole state may be described by the wave function $\ph(\bsr_1)\,\psi(\bsr_2)$, which is a complex-valued function of $(\bsr_1,\bsr_2)$.
Likewise one can take a state $\kappa(\bsr_1)\,\eta(\bsr_2)$ of the same two-particle system.
Then from the principle of superposition, we see that states of the form $\alpha\,\ph(\bsr_1)\,\psi(\bsr_2)+\beta\,\kappa(\bsr_1)\,\eta(\bsr_2)$ with $\alpha,\beta\in\bbC$ are also allowed.
Note that such a state is in general not written in the form $(\text{function of $\bsr_1$})\times(\text{function of $\bsr_2$})$.
Since any such superpositions are allowed, a general state of the two-particle system is described by a wave function $\Phi(\bsr_1,\bsr_2)$, which is an arbitrary complex-valued function\footnote{%
Recall that, with an arbitrary complete orthonormal system $\{\bsxi_\alpha\}_{\alpha=1,2,\ldots}$ of $\hilb$, one can express any function of $(\bsr_1,\bsr_2)$ as $\Phi(\bsr_1,\bsr_2)=\sum_{\alpha,\beta=1}^\infty c_{\alpha,\beta}\,\xi_\alpha(\bsr_1)\,\xi_\beta(\bsr_2)$.
} of $(\bsr_1,\bsr_2)\in\bbR^6$.

In exactly the same manner,  we see that a quantum mechanical state of a system of $N$ distinguishable particles is described by a wave function $\Phi(\roN)$, a complex-valued function of $\roN$, where $\bsr_j$ is the position of the $j$-th particle.
The wave function should satisfy the condition of square integrability:
\eq
\idrN\,|\Phi(\roN)|^2<\infty
\lb{si2}
\en

We again denote the whole wave function $\Phi(\roN)$ (where each of $\roN$ runs over the whole space) as $\bsPhi$.
The inner product of two wave functions $\bsPhi, \bsPsi$ is 
\eq
\sbkt{\bsPhi,\bsPsi}\coloneqq{}\idrN\,\{\Phi(\roN)\}^*\,\Psi(\roN),
\lb{ip}
\en
and the norm of a wave function $\bsPhi$ is $\snorm{\bsPhi}\coloneqq{}\sqrt{\sbkt{\bsPhi,\bsPhi}}$.

\paragraph*{Two indistinguishable particles}
Let us first consider a system of two identical particles.
A state of the system is described by a wave function $\Phi(\bsr_1,\bsr_2)$ satisfying the square integrability condition \rlb{si2}.
We shall see that the identity of particles imposes strict symmetry on the wave function.\footnote{%
This part has become longer than I first intended.  The reader who wishes to quickly move to the main topic may jump to \rlb{P12P21}.
}
%Since the two particles are identical, the wave function must satisfy strict symmetry.

Note that when we write $\bsr_1$, $\bsr_2$, we are assigning labels (i.e., 1 and 2) to the particles.
It is known, however, that, in quantum physics, two identical particles are intrinsically indistinguishable.
This, in particular, means that any physically meaningful observable must be invariant when the labels of the two particles are exchanged.\footnote{%
\label{fn:exchange}
It should be stressed that we are {\em not}\/ physically exchanging the two particles.
We are not making any changes to the physical state of the particles and merely changing the labels that we have assigned (rather arbitrarily).
}
For example, the operator\footnote{%
$\Lap_1=\frac{\partial^2}{\partial x_1^2}+\frac{\partial^2}{\partial y_1^2}+\frac{\partial^2}{\partial z_1^2}$ is the Laplacian corresponding to $\bsr_1=(x_1,y_1,z_1)$.
} $-\frac{\hbar^2}{2m}\Lap_1$, which represents the kinetic energy of the first particle, is not allowed since it certainly distinguishes between the labels of the particles.
The total kinetic energy $-\frac{\hbar^2}{2m}\Lap_1-\frac{\hbar^2}{2m}\Lap_2$ is an example of a legitimate observable.\footnote{%
We can, of course, measure the kinetic energy of a single particle when one of the two identical particles is in the measurement device and the other outside.
Such a measurement may be modeled (somewhat artificially) by the observable $-\frac{\hbar^2}{2m}\Lap_1\chi(\bsr_1)-\frac{\hbar^2}{2m}\Lap_2\chi(\bsr_2)$, where $\chi(\bsr)$ is 1 inside the measurement device and 0 outside.
(The domain of the operator consists of wave functions that vanish at the boundary of the device.)
Note that we are distinguishing the two particles by their positions, not by their labels.
}
In general, we require that any observable, when expressed in the Schr\"odinger representation, satisfies
\eq
A\Bigl(\bsr_1,\frac{d}{d\bsr_1};\bsr_2,\frac{d}{d\bsr_2}\Bigr)=A\Bigl(\bsr_2,\frac{d}{d\bsr_2};\bsr_1,\frac{d}{d\bsr_1}\Bigr).
\lb{A12=A21}
\en
We abbreviate this identity as $\hA_{12}=\hA_{21}$.

We next observe that any wave function $\Phi(\bsr_1,\bsr_2)$ is decomposed into the symmetric and the antisymmetric parts as
\eq
\Phi(\bsr_1,\bsr_2)=\Phi_+(\bsr_1,\bsr_2)+\Phi_-(\bsr_1,\bsr_2),
\lb{Phidec}
\en
where
\eq
\Phi_\pm(\bsr_1,\bsr_2)\coloneqq\frac{1}{2}\bigl\{\Phi(\bsr_1,\bsr_2)\pm\Phi(\bsr_2,\bsr_1)\bigr\}.
\en
These wave functions, of course, satisfy
\eq
\Phi_\pm(\bsr_1,\bsr_2)=\pm\Phi_\pm(\bsr_2,\bsr_1).
\en
Let us note that the two states $\bsPhi_+$ and  $\bsPhi_-$ are necessarily orthogonal (when nonzero).
This is seen by observing that
\eqa
\sbkt{\bsPhi_+,\bsPhi_-}&=\int d^3\bsr_1\,d^3\bsr_2\,\{\Phi_+(\bsr_1,\bsr_2)\}^*\,\Phi_-(\bsr_1,\bsr_2)
\nl&=\frac{1}{4}\int d^3\bsr_1\,d^3\bsr_2\bigl(|\Phi(\bsr_1,\bsr_2)|^2-|\Phi(\bsr_2,\bsr_1)|^2
\nl&\hspace{2.5cm}+\{\Phi(\bsr_2,\bsr_1)\}^*\,\Phi(\bsr_1,\bsr_2)-\{\Phi(\bsr_1,\bsr_2)\}^*\,\Phi(\bsr_2,\bsr_1)\bigr)
=0,
\ena
where we noted, e.g., that $\int d^3\bsr_1\,d^3\bsr_2|\Phi(\bsr_1,\bsr_2)|^2=\int d^3\bsr_1\,d^3\bsr_2|\Phi(\bsr_2,\bsr_1)|^2$ from the change of variables $\bsr_1\leftrightarrow\bsr_2$.

Let $\hA_{12}=\hA_{21}$ be any observable, and decompose its expectation value in the state $\bsPhi$ as
\eq
\sbkt{\bsPhi,\hA_{12}\,\bsPhi}=\sbkt{\bsPhi_+,\hA_{12}\,\bsPhi_+}+\sbkt{\bsPhi_-,\hA_{12}\,\bsPhi_-}+\sbkt{\bsPhi_+,\hA_{12}\,\bsPhi_-}+\sbkt{\bsPhi_-,\hA_{12}\,\bsPhi_+}.
\en
By using the symmetry \rlb{A12=A21}, we find that
\eqa
\sbkt{\bsPhi_+,\hA_{12}\,\bsPhi_-}&=\int d^3\bsr_1\,d^3\bsr_2\,\{\Phi_+(\bsr_1,\bsr_2)\}^*\,\hA_{12}\,\Phi_-(\bsr_1,\bsr_2)
\nl&=\frac{1}{4}\int d^3\bsr_1\,d^3\bsr_2\bigl(
\{\Phi(\bsr_1,\bsr_2)\}^*\,\hA_{12}\,\Phi(\bsr_1,\bsr_2)-\{\Phi(\bsr_2,\bsr_1)\}^*\,\hA_{12}\,\Phi(\bsr_2,\bsr_1)
\nl&\hspace{2.5cm}+\{\Phi(\bsr_2,\bsr_1)\}^*\,\hA_{12}\,\Phi(\bsr_1,\bsr_2)-\{\Phi(\bsr_1,\bsr_2)\}^*\,\hA_{12}\,\Phi(\bsr_2,\bsr_1)\bigr)
\nl&=\frac{1}{4}\int d^3\bsr_1\,d^3\bsr_2\bigl(
\{\Phi(\bsr_1,\bsr_2)\}^*\,\hA_{12}\,\Phi(\bsr_1,\bsr_2)-\{\Phi(\bsr_2,\bsr_1)\}^*\,\hA_{21}\,\Phi(\bsr_2,\bsr_1)
\nl&\hspace{2.5cm}+\{\Phi(\bsr_2,\bsr_1)\}^*\,\hA_{12}\,\Phi(\bsr_1,\bsr_2)-\{\Phi(\bsr_1,\bsr_2)\}^*\,\hA_{21}\,\Phi(\bsr_2,\bsr_1)\bigr)
\nl&=0,
\ena
and similarly that $\sbkt{\bsPhi_-,\hA_{12}\,\bsPhi_+}=0$.
We thus find for any physically meaningful observable $\hA_{12}$ that
\eq
\sbkt{\bsPhi,\hA_{12}\,\bsPhi}=\sbkt{\bsPhi_+,\hA_{12}\,\bsPhi_+}+\sbkt{\bsPhi_-,\hA_{12}\,\bsPhi_-},
\lb{P12P2100}
\en
which means that the state behaves as a statistical mixture rather than a pure state\footnote{One can say that $\bsPhi_+$ and $\bsPhi_-$ belong to distinct super-selection sectors.} unless $\bsPhi_+=0$ or $\bsPhi_-=0$.

It may be reasonable to postulate that any quantum system admits pure states.
By assuming that the wave function $\Phi(\bsr_1,\bsr_2)$ describes a pure state, \rlb{P12P2100} implies 
% for any two-particle wave function $\Phi(\bsr_1,\bsr_2)$, 
one must have either $\Phi_+(\bsr_1,\bsr_2)=0$ or $\Phi_-(\bsr_1,\bsr_2)=0$ for any $\bsr_1,\bsr_2$.
Equivalently, the wave function must satisfy the symmetry
\eq
\Phi(\bsr_1,\bsr_2)=\zeta\,\Phi(\bsr_2,\bsr_1),
\lb{P12P21}
\en
for any $\bsr_1,\bsr_2\in\bbR^3$, where $\zeta$ is either 1 or $-1$.
We found that the wave function is multiplied by the constant $\zeta$ when the labels of the two particles are exchanged.
(See footnote~\ref{fn:exchange} again.)

Note here that any wave function of a given system (of two identical particles) can be obtained by continuously modifying another wave function of the same system, with the constraint \rlb{P12P21} always being preserved.
Since $\zeta$ is either 1 or $-1$, the continuity implies it is constant.\footnote{To be rigorous, a wave function is not necessarily continuous.  But the conclusion is still valid because continuous wave functions are dense in the Hilbert space.}
In other words, $\zeta$ is common for any wave function of the system.

This observation suggests that particles in our universe are classified into one of the two classes with $\zeta=1$ or $\zeta=-1$.
This is indeed the case, and particles with $\zeta=1$ are known as bosons, and with $\zeta=-1$ as fermions.\footnote{%
In the real world, any (elementary) particle is associated with spin, whose spin quantum number is an integer for bosons and a half-odd-integer for fermions.
Here we are trying to justify (or, at least, to motivate) the symmetry requirement \rlb{P12P21} without invoking spins.

Let us note that although we are focusing on three-dimensional systems, the dimension does not play any essential role in the present discussion.
The conclusion that a particle is either a boson or a fermion is legitimate in any dimension.
It is known that ``particles'' (or quasi-particles, to be precise) called anyons, which are neither bosons nor fermions, appear in certain two-dimensional systems.
But these ``particles'' are never described by a wave function like $\Phi(\bsr_1,\bsr_2)$.
}
Electrons are fermions, and atoms may be fermions or bosons. 
In the rest of the present note, we always understand that $\zeta$ is fixed to either 1 or $-1$ depending on the kind of particles that we are treating.

\paragraph*{$N$ indistinguishable particles}
Let us extend this consideration to a system of $N$ identical particles.
We take a wave function $\Phi(\roN)$ which satisfies the square integrability condition \rlb{si2}.
To take into account the symmetry with respect to the change of the labels of the particles, we further assume that the wave function satisfies
\eq
\Phi(\bsr_1,\bsr_2,\ldots,\bsr_N)=\zeta^P\,\Phi(\bsr_{P(1)},\bsr_{P(2)},\ldots,\bsr_{P(N)}),
\lb{symN}
\en
for an arbitrary permutation $P$ of $\{1,2,\ldots,N\}$.\footnote{%
This postulate may be partially justified as follows.
We treat the case with $N=3$ for simplicity.

Consider a three-particle state, described by the wave function $\Phi(\bsr_1,\bsr_2,\bsr_3)$, in which two of the three particles are confined in a finite region $V\subset\bbR^3$, and the other particle is confined in a sufficiently separated finite region $V'$.
It is then plausible that, when focusing on the two particles in the region $V$, one does not have to take into account the presence of the third particle in $V'$.
Then, if we only consider the case $\bsr_1,\bsr_2\in V$ and $\bsr_3\in V'$, the three-particle wave function should satisfy $\Phi(\bsr_1,\bsr_2,\bsr_3)=\zeta\,\Phi(\bsr_2,\bsr_1,\bsr_3)$, reflecting the symmetry \rlb{P12P21} of two-particle wave functions.
Similarly, we have $\Phi(\bsr_1,\bsr_2,\bsr_3)=\zeta\,\Phi(\bsr_1,\bsr_3,\bsr_2)$ when $\bsr_2,\bsr_3\in V$ and $\bsr_1\in V'$, and $\Phi(\bsr_1,\bsr_2,\bsr_3)=\zeta\,\Phi(\bsr_3,\bsr_2,\bsr_1)$ when $\bsr_3,\bsr_1\in V$ and $\bsr_2\in V'$.

Suppose that one continuously modifies the wave function to get a state in which the three particles are not necessarily confined in separate regions.
In other words, one moves and modifies the two regions $V$ and $V'$ so that they overlap.
It is reasonable to assume that the above three symmetry relations are all preserved during this modification.
We then get the desired symmetry \rlb{symN} for any $\bsr_1,\bsr_2,\bsr_3$.
}
%More precisely a permutation $P(j)$ is a function of $j=1,2,\ldots,N$ such that $P(j)\ne P(j')$ if $j\ne j'$.
%Note that there are exactly $N!$ distinct permutations.
%We set $\zeta^P=1$ when $\zeta=1$, and $\zeta^P=(-1)^P$, the parity of the permutation $P$, when $\zeta=-1$.
We here set
\eq
\zeta^P=\begin{cases}
1&\text{for bosons with $\zeta=1$},\\
(-1)^P&\text{for fermions with $\zeta=-1$},
\end{cases}
\en
where $(-1)^P=\pm1$ denote the parity of the permutation $P$.

The inner product and the norm of these states are defined exactly as in the distinguishable case.  See \rlb{ip}.
For any $N=1,2,\ldots$, we denote by $\calH_N$ the space consisting of all such wave functions.
We call $\calH_N$ the $N$-particle Hilbert space.
We also identify the 0-particle Hilbert space $\calH_0$ with $\bbC$, the set of complex numbers.

\paragraph*{Purpose of the present note}
What we have briefly described above provides a complete quantum mechanical description of a system of multiple identical particles.\footnote{%
The formalism can be automatically extended to include spin degrees of freedom.}
One can formulate various advanced problems in this language and perform calculations.
It is however awkward in some sense to first assing the labels $1,2,\ldots,N$ to particles (which cannot be labeled indeed), write down the wave function, and finally impose additional symmetry constraint \rlb{symN} to take into account the indistinguishability.
This procedure sometimes makes practical calculations complicated.

By using the creation and annihilation operators and the Fock representation, one can describe quantum states of indistinguishable particles and operators acting on them without ever assigning labels to the particles.
This formulation is not only theoretically beautiful but is also useful in practice.
It should be stressed, however, that this formulation is strictly equivalent to the formulation in terms of wave functions described above; there is nothing new physically.
One might be tempted to imagine that the formulation in terms of the creation/annihilation operators contains some novel physics compared with the wave function formulation, especially because the former is also known by the name ``second quantization''.
But this is a misunderstanding.
See the beginning of section~\ref{ss:SQ}.

%%%%%%%%%%%%%%%%%%%%%%%%%%%%%%%%%%%%%%
\section{Creation and annihilation operators}
\label{ss:CA}
Let us discuss the creation and annihilation operators, which will play a central role in the present note.
% ``second quantization'' formalism.
%In order to give a connection between the wave function formalism in section~\ref{ss:WF} and the occupation number representation in section~\ref{ss:SQ}, we here construct and study creation and annihilation operators within the language of wave functions.

\index{creation operator}
\paragraph*{Creation operator}
For any $\bspsi\in\hilb$ and $N=1,2,\ldots$, we wish to define the creation operator $\tCd(\bspsi):\calHwf_{N-1}\to\calHwf_N$, which ``adds'' the state $\bspsi$ to an arbitrary $N-1$ particle state $\bsPhi$ to generate a new $N$ particle state $\tCd(\bspsi)\bsPhi$.

When $N=2$ a natural definition of such an operator is
\eq
(\tCd(\bspsi)\bsph)(\bsr_1,\bsr_2)=\frac{1}{\sqrt{2}}\bigl\{
\psi(\bsr_1)\ph(\bsr_2)+\zeta\,\psi(\bsr_2)\ph(\bsr_1)
\bigr\},
\lb{CA1}
\en
where $\bsph\in\hilb$ is an arbitrary single-particle state.\footnote{%
We write the wave function representation $\Phi(\roN)$ of a state $\bsPhi\in\calH_N$ also as $(\bsPhi)(\roN)$.
}
Note that $\psi(\bsr_1)\ph(\bsr_2)+\zeta\,\psi(\bsr_2)\ph(\bsr_1)$ is the only symmetric or antisymmetric two particle wave functions that one can generate from $\psi(\bsr)$ and $\ph(\bsr)$.
The norm of the new state is readily found to be
\eqa
\snorm{\tCd(\bspsi)\bsph}^2
&=\frac{1}{2}
\int d^3\bsr_1\,d^3\bsr_2\,
\bigl\{\psi(\bsr_1)\ph(\bsr_2)+\zeta\,\psi(\bsr_2)\ph(\bsr_1)\bigr\}^*\,
\bigl\{\psi(\bsr_1)\ph(\bsr_2)+\zeta\,\psi(\bsr_2)\ph(\bsr_1)\bigr\}
\nl&=\snorm{\bspsi}^2\,\snorm{\bsph}^2+\zeta\,|\sbkt{\bspsi,\bsph}|^2.
\lb{CA2}
\ena
Suppose that $\bspsi$ and $\bsph$ are normalized, i.e., $\snorm{\bspsi}=\snorm{\bsph}=1$.
We see from \rlb{CA2} that  the new state $\tCd(\bspsi)\bsph$ is normalized only when $\sbkt{\bspsi,\bsph}=0$.
Note also that, for fermions with $\zeta=-1$, we have $\tCd(\bspsi)\bsph=0$ when $\bspsi=e^{i\theta}\bsph$.
This is a mathematical expression of the Pauli exclusion principle, which inhibits two fermions to occupy the same single-particle state.

Let us consider the most straightforward extension of  \rlb{CA1} to general $N$.
For an arbitrary $\bsPhi\in\calHwf_{N-1}$, we define
\eq
(\tCd(\bspsi)\bsPhi)(\bsr_1,\ldots,\bsr_N)\coloneqq{}\frac{1}{\sqrt{N}}\sum_{j=1}^N
\zeta^{j-1}\,\psi(\bsr_j)\,\Phi(\bsr_1,\ldots,\breve{\bsr}_j,\ldots,\bsr_N),
\lb{CA3}
\en
for any $\bsr_1,\ldots,\bsr_N\in\Rt$.
Here and in what follows we denote by $\bsr_1,\ldots,\breve{\bsr}_j,\ldots,\bsr_N$ the sequence without $\bsr_j$, i.e., $\bsr_1,\ldots,\bsr_{j-1},\bsr_{j+1},\ldots,\bsr_N$.
Thus  \rlb{CA3} with $N=3$ reads
\eq
(\tCd(\bspsi)\bsPhi)(\bsr_1,\bsr_2,\bsr_3)=\frac{1}{\sqrt{3}}\bigl\{
\psi(\bsr_1)\,\Phi(\bsr_2,\bsr_3)+\zeta\,\psi(\bsr_2)\,\Phi(\bsr_1,\bsr_3)+\psi(\bsr_3)\,\Phi(\bsr_1,\bsr_2)
\bigr\}.
\lb{CA3B}
\en
%The reader is encouraged to write down \rlb{CA3} for $N=4$.
Note that the right-hand side of \rlb{CA3} or \rlb{CA3B} satisfy the symmetry \rlb{symN}.
For $N=1$ and $\bsPhi=1\in\calHwf_0$, the definition \rlb{CA3} gives
\eq
(\tCd(\bspsi)1)(\bsr)=\psi(\bsr).
\lb{CA4}
\en
It is clear from the definition \rlb{CA3} that  the creation operator $\tCd(\bspsi)$ is linear in $\bspsi$, i.e., 
\eq
\tCd\Bigl(\sum_{\ell=1}^nc_\ell\,\bspsi_\ell\Bigr)=\sum_{\ell=1}^nc_\ell\,\tCd(\bspsi_\ell),
\lb{CA4B}
\en
for any states $\bspsi_1,\ldots,\bspsi_n\in\hilb$ and coefficients $c_1,\ldots,c_n\in\bbC$.

\index{annihilation operator}
\paragraph*{Annihilation operator}
We define the annihilation operator $\tC(\bspsi):\calHwf_N\to\calHwf_{N-1}$ for any $N=1,2,\ldots$ by $\tC(\bspsi)\coloneqq{}\{\tCd(\bspsi)\}^\dagger$.
It should be the operator that ``removes'' the state $\bspsi$ from an arbitrary $N$ particle state $\bsPhi$ to generate a new $N-1$ particle state $\tC(\bspsi)\bsPhi$.
Let us determine the action of $\tC(\bspsi)$ from the identity
\eq
\sbkt{\tCd(\bspsi)\bsXi,\bsPhi}=\sbkt{\bsXi,\tC(\bspsi)\bsPhi},
\lb{CA5}
\en
for arbitrary $\bsXi\in\calHwf_{N-1}$ and $\bsPhi\in\calHwf_N$ with $N=1,2,\ldots$.
From \rlb{CA3}, we find
%\begingroup\allowdisplaybreaks
\eqa
\langle\tCd&(\bspsi)\bsXi,\bsPhi\rangle
\nl&
=
\idrN\,\frac{1}{\sqrt{N}}\sum_{j=1}^N\zeta^{j-1}\bigl\{\psi(\bsr_j)\,\Xi(\bsr_1,\ldots,\breve{\bsr}_j,\ldots,\bsr_N)
\bigr\}^*\,
\Phi(\bsr_1,\ldots,\bsr_N)\notag
\intertext{%
Let us fix $j$, and write the sequence $(\bsr_1,\ldots,\breve{\bsr}_j,\ldots,\bsr_N)$ as $(\bss_1,\ldots,\bss_{N-1})$ (or, in other words, set $\bss_i=\bsr_i$ for $i<j$ and $\bss_{i-1}=\bsr_i$ for $i>j$).
Then we have $\Phi(\bsr_1,\ldots,\bsr_N)=$\newline$=\Phi(\bss_1,\ldots,\bss_{j-1},\bsr_j,\bss_j,\ldots,\bss_{N-1})=\zeta^{j-1}\,\Phi(\bsr_j,\bss_1,\ldots,\bss_{N-1})$, where we used the symmetry \rlb{symN}.
Since $(\zeta^{j-1})^2=1$, we now have}
&=
\int d^3\bss_1\ldots d^3\bss_{N-1}\,\frac{1}{\sqrt{N}}\sum_{j=1}^N\int d^3\bsr_j\,
\bigl\{\psi(\bsr_j)\,\Xi(\bss_1,\ldots,\bss_{N-1})
\bigr\}^*\,
\Phi(\bsr_j,\bss_1,\ldots,\bss_{N-1})\notag
\intertext{%
By rewriting $\bsr_j$ as $\bsq$, we see that the summands with different $j$ produce exactly the same expression, and hence
}
&=\int d^3\bss_1\ldots d^3\bss_{N-1}\,\{\Xi(\bss_1,\ldots,\bss_{N-1})\}^*\,\sqrt{N}
\int d^3\bsq\,\{\psi(\bsq)\}^*\,\Phi(\bsq,\bss_1,\ldots,\bss_{N-1}).
\lb{CA6}
\ena
%\endgroup
By comparing this expression with the right-hand side of \rlb{CA5}, we find for any $\bsPhi\in\calHwf_N$ with $N=1,2,\ldots$ that
\eq
(\tC(\bspsi)\bsPhi)(\bsr_1,\ldots,\bsr_{N-1})=\sqrt{N}\int d^3\bsq\,\{\psi(\bsq)\}^*\,\Phi(\bsq,\bsr_1,\ldots,\bsr_{N-1}),
\lb{CA7}
\en
which is the desired characterization of the annihilation operator.
The wave function \newline$(\tC(\bspsi)\bsPhi)(\bsr_1,\ldots,\bsr_{N-1})$ clearly satisfies the desired symmetry \rlb{symN}.
We also define
\eq
\tC(\bspsi)c_0=0,
\lb{CA7B}
\en
for any $c_0\in\calHwf_0\cong\bbC$ since $\tCd(\bspsi)\bsPhi\not\in\calHwf_0$ for all possible $\bsPhi$.
Note that  the annihilation operator $\tC(\bspsi)$ is antilinear in $\bspsi$, i.e., 
\eq
\tC\Bigl(\sum_{\ell=1}^nc_\ell\,\bspsi_\ell\Bigr)=\sum_{\ell=1}^n(c_\ell)^*\,\tC(\bspsi_\ell),
\lb{CA4BB}
\en
for any states $\bspsi_1,\ldots,\bspsi_n\in\hilb$ and coefficients $c_1,\ldots,c_n\in\bbC$.
This is clear from the expression \rlb{CA7}, but follows readily from \rlb{CA4B}.

\index{anticommutation relation}
\paragraph*{Anticommutation relations}
We shall derive the (anti)commutation relations \rlb{CA10}, \rlb{CA10B}, and \rlb{CA15}, which provide the essential characterization of the creation and annihilation operators.
Let $\bsph,\bspsi\in\hilb$.
For any $\bsXi\in\calHwf_N$ wit $N=2,3,\ldots$, we find by using \rlb{CA7} twice that
\eqa
(\tC(\bsph)\tC(\bspsi)\bsXi)(\bsr_1,\ldots,\bsr_{N-2})
&=\sqrt{N-1}\int d^3\bsq\{\ph(\bsq)\}^*(\tC(\bspsi)\bsXi)(\bsq,\bsr_1,\ldots,\bsr_{N-2})
\nl&=\sqrt{N(N-1)}\int d^3\bsq\,d^3\bsq'\{\ph(\bsq)\,\psi(\bsq')\}^*\,\Xi(\bsq',\bsq,\bsr_1,\ldots,\bsr_{N-2}),
\lb{CA8}
\ena
and
\eq
(\tC(\bspsi)\tC(\bsph)\bsXi)(\bsr_1,\ldots,\bsr_{N-2})=\sqrt{N(N-1)}\int d^3\bsq\,d^3\bsq'\{\psi(\bsq')\,\ph(\bsq)\}^*\,\Xi(\bsq,\bsq',\bsr_1,\ldots,\bsr_{N-2}).
\lb{CA9}
\en
Since $\Xi(\bsq,\bsq',\bsr_1,\ldots,\bsr_{N-2})=\zeta\,\Xi(\bsq',\bsq,\bsr_1,\ldots,\bsr_{N-2})$, we find
\eq
\tC(\bsph)\tC(\bspsi)\bsXi=\zeta\,\tC(\bspsi)\tC(\bsph)\bsXi,
\lb{CA9B}
\en
for any $\bsXi\in\calHwf_N$ wit $N=2,3,\ldots$.
Let us define $[\hA,\hB]_{-\zeta}\coloneqq{}\hA\hB-\zeta\,\hB\hA$ for any operators $\hA$ and $\hB$.
For bosons, where $\zeta=1$, this defines the standard commutator, and for fermions, where $\zeta=-1$, this defines the anticommutator.
Recalling that $\bsXi$ is arbitrary in \rlb{CA9B}, we find  the following (anti)commutation relations for the annihilation operators
\eq
[\tC(\bsph),\tC(\bspsi)]_{-\zeta}=0,
\lb{CA10}
\en
for any $\bsph,\bspsi\in\hilb$.
By taking the adjoint, we  find for the creation operators that
\eq
[\tCd(\bsph),\tCd(\bspsi)]_{-\zeta}=0.
\lb{CA10B}
\en
For fermions, where we set $\zeta=-1$, one finds by setting $\bsph=\bspsi$ in \rlb{CA10} and  \rlb{CA10B} that
\eq
\{\tC(\bsph)\}^2=\{\tCd(\bsph)\}^2=0\quad(\text{only for fermions!})
\lb{CA10C}
\en
This is another mathematical expression of the Pauli exclusion principle.

To evaluate the (anti)commutator $[\tC(\bsph),\tCd(\bspsi)]_{-\zeta}$ is interesting but a little complicated.
We encourage the reader to explicitly write down the following derivation in the case with $N=2$.\footnote{%
One of the undergraduate students in our group indeed worked this out on the blackboard.
The demonstration turned out to be quite useful.
}
Take an arbitrary $\bsXi\in\calHwf_N$ with $N=1,2,\ldots$.
First it is easy to see from \rlb{CA3} and \rlb{CA7} that
\eqa
(\tCd(\bspsi)\tC(\bsph)\bsXi)(\bsr_1,\ldots,\bsr_N)
&=\frac{1}{\sqrt{N}}\sum_{j=1}^N\zeta^{j-1}\psi(\bsr_j)(\tC(\bsph)\bsXi)(\bsr_1,\ldots,\breve{\bsr}_j,\ldots,\bsr_N)
\nl&=\sum_{j=1}^N\zeta^{j-1}\psi(\bsr_j)\int d^3\bsq\,\{\ph(\bsq)\}^*\,\Xi(\bsq,\bsr_1,\ldots,\breve{\bsr}_j,\ldots,\bsr_N).
\lb{CA11}
\ena
Next we simply use \rlb{CA7} to observe that
\eq
(\tC(\bsph)\tCd(\bspsi)\bsXi)(\bsr_1,\ldots,\bsr_N)
=\sqrt{N+1}\int d^3\bsq\,\{\ph(\bsq)\}^*(\tCd(\bspsi)\bsXi)(\bsq,\bsr_1,\ldots,\bsr_N).
\lb{CA11B}
\en
To rewrite the right-hand side, we note that \rlb{CA3} implies
\eqa
(\tCd&(\bspsi)\bsXi)(\bsq,\bsr_1,\ldots,\bsr_N)
\nl&=\frac{1}{\sqrt{N+1}}\Bigl\{\psi(\bsq)\,\Xi(\bsr_1,\ldots,\bsr_N)
+\sum_{j=1}^N\zeta^j\psi(\bsr_j)\,\Xi(\bsq,\bsr_1,\ldots,\breve{\bsr}_{j},\ldots,\bsr_{N})\Bigr\}.
\lb{CA12}
\ena
Note that $j$ in this expression corresponds to $j-1$ in \rlb{CA3}.
Substituting this back to \rlb{CA11B}, we get
\eqa
(\tC&(\bsph)\tCd(\bspsi)\bsXi)(\bsr_1,\ldots,\bsr_N)
\nl&=
\int d^3\bsq\,\{\ph(\bsq)\}^*\Bigl\{
\psi(\bsq)\,\Xi(\bsr_1,\ldots,\bsr_N)
+\sum_{j=1}^N\zeta^j\psi(\bsr_j)\,\Xi(\bsq,\bsr_1,\ldots,\breve{\bsr}_{j},\ldots,\bsr_{N})\Bigr\}.
\lb{CA13}
\ena
From \rlb{CA11} and  \rlb{CA13}, we finally obtain
\eq
\Bigl(\tC(\bsph)\tCd(\bspsi)\bsXi-\zeta\,\tCd(\bspsi)\tC(\bsph)\bsXi\Bigr)(\bsr_1,\ldots,\bsr_N)=\sbkt{\bsph,\bspsi}\,\Xi(\bsr_1,\ldots,\bsr_N),
\lb{CA14}
\en
for any $\bsXi$.
This leads to the (anti)commutation relation
\eq
[\tC(\bsph),\tCd(\bspsi)]_{-\zeta}=\sbkt{\bsph,\bspsi},
\lb{CA15}
\en
for any $\bsph,\bspsi\in\hilb$.

\paragraph*{Examples of annihilation and creation operators}
Take an arbitrary complete orthonormal system $\{\bsxi_\alpha\}_{\alpha=1,2,\ldots}$ of the single particle Hilbert space $\hilb$.
%A useful example is to take $\bsxi_\alpha$ as eigenstates of a single-particle Hamiltonian $\hH_1$, i.e., $\hH_1\bsxi_\alpha=\epsilon_\alpha\bsxi_\alpha$.
We then define
\eq
\ha_\alpha\coloneqq{}\tC(\bsxi_\alpha),\quad\had_\alpha\coloneqq{}\tCd(\bsxi_\alpha),
\lb{haa}
\en
which are annihilation and creation operators of the state $\bsxi_\alpha$.
We see immediately from  \rlb{CA10}, \rlb{CA10B}, and \rlb{CA15} that
\eq
[\ha_\alpha,\ha_\beta]_{-\zeta}=[\had_\alpha,\had_\beta]_{-\zeta}=0,\quad
[\ha_\alpha,\had_\beta]_{-\zeta}=\delta_{\alpha,\beta},
\lb{aaab}
\en
for any $\alpha,\beta=1,2,\ldots$.
It is remarkable that, for bosons, these are nothing but the commutation relations for the raising and lowering operators in a system of harmonic oscillators.

For an arbitrary $\bsx\in\Rt$, consider a state $\eta^{\bsx}(\bsr)\coloneqq{}\delta(\bsr-\bsx)$, in which the particle is completely localized at $\bsx$.\footnote{%
In physics literature, including Feynman's ``Statistical Mechanics'' \cite{Feynman}, the state $\bseta^{\bsx}$ is denoted as $\ket{\bsx}$.
}
Recall that the inner product of two such states is 
\eq
\sbkt{\bseta^{\bsx},\bseta^{\bsy}}=\idr\,\delta(\bsr-\bsx)\,\delta(\bsr-\bsy)=\delta(\bsx-\bsy).
\lb{ddin}
\en
Consider the annihilation and creation operators corresponding the state $\bseta^{\bsx}$, i.e., 
\eq
\hpsi(\bsx)\coloneqq{}\ha(\bseta^{\bsx}), \quad \hpsid(\bsx)\coloneqq{}\had(\bseta^{\bsx}),
\lb{hpx}
\en
which annihilates or creates a particle at $\bsx$.\footnote{%
In Feynman's notation \cite{Feynman}, $\hpsi(\bsx)$ and $\hpsid(\bsx)$ become $\ha(\bsx)$ and $\had(\bsx)$.
Our notation is also standard.

The careful reader might have noticed that $\bseta^{\bsx}$ does not satisfy the square integrability condition \rlb{si}, and is not a proper single-particle state.
We nevertheless define  the operators $\hpsi(\bsx)$ and  $\hpsid(\bsx)$ (rather formally), since they turn out to be useful.
}
From  \rlb{CA10}, \rlb{CA10B}, \rlb{CA15}, and \rlb{ddin}, we see that they satisfy the (anti)commutation relations
\eq
[\hpsi(\bsx),\hpsi(\bsy)]_{-\zeta}=[\hpsid(\bsx),\hpsid(\bsy)]_{-\zeta}=0,\quad
[\hpsi(\bsx),\hpsid(\bsy)]_{-\zeta}=\delta(\bsx-\bsy),
\lb{pp}
\en
for any $\bsx,\bsy\in\Rt$.

%The careful reader might have noticed that $\bseta^{\bsx}$ does not satisfy the square integrability condition \rlb{si}, and hence cannot be regarded as a proper single-particle state.
%Consequently the operators $\hpsi(\bsx)$ and  $\hpsid(\bsx)$ are only formally defined 

Let us see how the operator $\hpsid(\bsx)$ is related to the standard operators $\had(\bsph)$ with $\bsph\in\hilb$.
%{\bf Probably I don't need the first relation.}
First, take an arbitrary single-particle state  $\bsph\in\hilb$.
Since $\ph(\bsr)=\int d^3\bsx\,\ph(\bsx)\,\delta(\bsr-\bsx)$, we see that $\bsph=\int d^3\bsx\,\ph(\bsx)\,\bseta^{\bsx}$.
Then from the linearity \rlb{CA4B} of the creation operator, we have
\eq
\had(\bsph)=\int d^3\bsx\,\ph(\bsx)\,\hpsid(\bsx).
\lb{adpd}
\en
Next take an arbitrary complete orthonormal system $\{\bsxi_\alpha\}_{\alpha=1,2,\ldots}$ of $\hilb$, and note that\footnote{%
This is the condition of completeness of $\{\bsxi_\alpha\}_{\alpha=1,2,\ldots}$.
To see this relation, take the inner product of the state  $\xi_\alpha(\bsr)$ and the assumed expansion $\delta(\bsr-\bsx)=\sum_{\beta=1}^\infty c_\beta\,\xi_\beta(\bsr)$ to find that $c_\alpha= \{\xi_\alpha(\bsx)\}^*$.
} $\delta(\bsr-\bsx)=\sum_{\alpha=1}^\infty \{\xi_\alpha(\bsx)\}^*\xi_\alpha(\bsr)$, and hence $\bseta^{\bsx}=\sum_{\alpha=1}^\infty \{\xi_\alpha(\bsx)\}^*\bsxi_\alpha$.
Again by using the linearity, we find
\eq
\hpsid(\bsx)=\sum_{\alpha=1}^\infty \{\xi_\alpha(\bsx)\}^*\,\had(\bsxi_\alpha),
\lb{pdad}
\en
which will turn out to be useful later.

For an arbitrary wave number vector $\bsk\in\Rt$, define the standard plane-wave state\footnote{%
This state does not satisfy the square integrability condition \rlb{si}.
} by
\eq
u^{\bsk}(\bsr)=(2\pi)^{-3/2}\,e^{i\bsk\cdot\bsr}.
\lb{ukr}
\en
From the standard expression of the delta function
\eq
\delta(\bsz)=\int\frac{d^3\bsw}{(2\pi)^3}\,e^{i\bsw\cdot\bsz}\quad\text{for $\bsz\in\Rt$},
\lb{delta}
\en
one finds that the inner product of two plane wave states is given by
\eq
\sbkt{\bsu^{\bsk},\bsu^{\bsk'}}=\delta(\bsk-\bsk'),
\en
for any $\bsk,\bsk'\in\Rt$.
We then define the corresponding annihilation and creation operators as\footnote{%
In Feynman's notation \cite{Feynman}, $\hak(\bsk)$ and $\hakd(\bsk)$ become $\ha(\bsk)$ and $\had(\bsk)$.
In this case, his notation seems to be standard.
}
\eq
\hak(\bsk)\coloneqq{}\ha(\bsu^{\bsk}),\quad\hakd(\bsk)\coloneqq{}\had(\bsu^{\bsk}),
\en
which satisfies the (anti)commutation relations
\eq
[\hak(\bsk),\hak(\bsk')]_{-\zeta}=[\hakd(\bsk),\hakd(\bsk')]_{-\zeta}=0,\quad
[\hak(\bsk),\hakd(\bsk')]_{-\zeta}=\delta(\bsk-\bsk'),
\en
for any $\bsk,\bsk'\in\Rt$.
From \rlb{ukr} and \rlb{delta}, we see that
\eq
\delta(\bsr-\bsx)=\int\frac{d^3\bsk}{(2\pi)^{3/2}}\,e^{-i\bsk\cdot\bsx}\,u^{\bsk}(\bsr).
\en
This equality, with the linearity \rlb{CA4B}, implies
\eq
\hpsid(\bsx)=\int\frac{d^3\bsk}{(2\pi)^{3/2}}\,e^{-i\bsk\cdot\bsx}\,\hakd(\bsk).
\lb{psdad}
\en

\section{The Fock space representation}
\label{ss:SQ}
We now discuss the description, which is often called the Fock space representation, of many-particle quantum mechanics in terms of the creation and annihilation operators.
This formalism is also known by the name ``second quantization'' formalism.
But one should note that this formalism is nothing more than a clever way of rewriting the standard quantum mechanics based on wave functions.
% it is strictly equivalent to the standard (or the ``first quantized'') formalism based on wave functions.
The unfortunate and misleading name ``second quantization'' simply reflects the (constructive) confusion in early history.

For convenience let us summarize the (anti)commutation relations \rlb{CA10}, \rlb{CA10B}, and \rlb{CA15} obeyed by the creation and annihilation operators:
\eq
[\tC(\bsph),\tC(\bspsi)]_{-\zeta}=[\tCd(\bsph),\tCd(\bspsi)]_{-\zeta}=0,\quad
[\tC(\bsph),\tCd(\bspsi)]_{-\zeta}=\sbkt{\bsph,\bspsi}
\quad\text{for any $\bsph,\bspsi\in\hilb$}
\lb{comm}
\en

\paragraph*{The representation of states}
We start by introducing a new description of the states in $\calH_N$ with $N=1,2,\ldots$.
First, let us take $1\in\calH_0\cong\bbC$, which is a ``state'' without any particles, and write it as $\vac$.
Here ``vac'' stands for the ``vacuum''.
From \rlb{CA7B}, the vacuum state satisfies
\eq
\ha(\bsph)\vac=0,
\lb{av0}
\en
for any $\bsph\in\hilb$.
This relation is repeatedly used in most applications of the Fock space representation.

For an arbitrary $\bsph\in\hilb$, we see from \rlb{CA4} that $\had(\bsph)\vac$ is the state $\bsph$ itself.
Then for an arbitrary $\bspsi\in\hilb$ one sees that $\had(\bspsi)\,\had(\bsph)\vac$ represents the state \rlb{CA1} with two particles.
Although not all states in $\calH_2$ is written in the form  \rlb{CA1}, one can recover any state in $\calH_2$ by considering arbitrary superpositions of two-particle states of the form \rlb{CA1}.

This consideration can be readily generalized to an arbitrary $N=1,2,\ldots$.
Let $\bsph_1,\ldots,\bsph_N\in\hilb$ be arbitrary states.
Then $\had(\bsph_1)\cdots\had(\bsph_N)\vac$ (if non-vanishing) is a state in $\calH_N$.
The $N$ particle Hilbert space $\calH_N$ is recovered by considering arbitrary superpositions of states of the form $\had(\bsph_1)\cdots\had(\bsph_N)\vac$.

Take arbitrary $\bsph_1,\ldots,\bsph_N,\bspsi_1,\ldots,\bspsi_N\in\hilb$, and consider two states
\eq
\ket{\Phi}=\had(\bsph_1)\cdots\had(\bsph_N)\vac,\quad\ket{\Psi}=\had(\bspsi_1)\cdots\had(\bspsi_N)\vac.
\lb{PhiPsi1}
\en
To state the following fundamental and useful result,  we define, for any $N\times N$ matrix $\sfA=(a_{i,j})_{i,j=1,\ldots,N}$,
\eq
|\sfA|_\zeta\coloneqq{}\sum_P\zeta^P\,a_{1,P(1)}\,a_{2,P(2)}\cdots a_{N,P(N)},
\en
where the sum runs over all $N!$ permutations of $\{1,2,\ldots,N\}$.
Thus $|\sfA|_-$ is the standard determinant, and $|\sfA|_+$ is the quantity known as permanent.

\begin{T}
The inner product of the two states defined in \rlb{PhiPsi1} is given by
\eq
\braket{\Phi}{\Psi}=\dmat{
\sbkt{\bsph_1,\bspsi_1}&\sbkt{\bsph_1,\bspsi_2}&\cdots&\sbkt{\bsph_1,\bspsi_N}\\
\sbkt{\bsph_2,\bspsi_1}&\sbkt{\bsph_2,\bspsi_2}&\cdots&\sbkt{\bsph_2,\bspsi_N}\\
\vdots&\vdots&\ddots&\vdots\\
\sbkt{\bsph_N,\bspsi_1}&\sbkt{\bsph_N,\bspsi_2}&\cdots&\sbkt{\bsph_N,\bspsi_N}
}_\zeta.
\lb{PhiPsi}
\en
\end{T}
\nproof{Proof}
The relation  can be shown by noting that
\eq
\braket{\Phi}{\Psi}=\vacb\ha(\bsph_N)\cdots\ha(\bsph_1)\had(\bspsi_1)\cdots\had(\bspsi_N)\vac,
\en
and repeatedly using $\ha(\bsph)\had(\bspsi)=\zeta\,\had(\bspsi)\ha(\bsph)+\sbkt{\bsph,\bspsi}$, which is the first of the (anti)commutation relations \rlb{comm}, and $\ha(\bsph)\vac=0$.
First, observe that
\eq
\ha(\bsph_1)\had(\bspsi_1)\cdots\had(\bspsi_N)\vac
=\sum_{j=1}^N\zeta^{j-1}\,\sbkt{\bsph_1,\bspsi_j}\,\had(\bspsi_1)\cdots\had(\bspsi_{j-1})\had(\bspsi_{j+1}),\cdots\had(\bspsi_N)\vac
\lb{adadaav}
\en
where $\had(\bspsi_j)$ is missing in the summand.
We can repeat this process in \newline$\ha(\bsph_N)\cdots\ha(\bsph_1)\had(\bspsi_1)\cdots\had(\bspsi_N)$ until all $\had$ and $\ha$ are gone.
The reader should work out the case with $N=3$ explicitly to check that the desired \rlb{PhiPsi} follows.
For general $N$, we clearly see
\eq
\ha(\bsph_N)\cdots\ha(\bsph_1)\had(\bspsi_1)\cdots\had(\bspsi_N)\vac
=\sum_P\eta(P)\prod_{j=1}^N\sbkt{\bsph_j,\bspsi_{P(j)}}\vac,
\en
and hence
\eq
\braket{\Phi}{\Psi}=\sum_P\eta(P)\prod_{j=1}^N\sbkt{\bsph_j,\bspsi_{P(j)}},
\lb{PP1}
\en
where $P$ is summed over all permutations of $\{1,\ldots,N\}$, and $\eta(P)=\pm1$ is a certain (still undetermined) sign factor which depends only on $P$.
For bosons with $\zeta=1$, it is clear that $\eta(P)=1$, and we are done.
We still need to show $\eta(P)=(-1)^P$ for fermions with $\eta=-1$.
This can be done by examining the sign appearing in relations like \rlb{adadaav}, but there is an easier trick.
Fix a permutation $P_0$, and note that the anticommutation relation \rlb{comm} implies
\eq
\braket{\Phi}{\Psi}=(-1)^{P_0}\,\vacb\ha(\bsph_N)\cdots\ha(\bsph_1)\had(\bspsi_{P_0(1)})\cdots\had(\bspsi_{P_0(N)})\vac.
\en
Exactly as before the right-hand side can be rewritten as in \rlb{PP1}, but we immediately see (without any calculations) that the result contains the term $(-1)^{P_0}\prod_{j=1}^N\sbkt{\bsph_j,\bspsi_{P_0(j)}}$, which has the desired sign factor.
Since all the states $\bsph_1,\ldots,\bsph_N$, $\bspsi_1,\ldots,\bspsi_N$ are arbitrary, and $P_0$ is also arbitrary, we have shown that $\eta(P)=(-1)^P$.~\qedm

\bigskip
Let us make a short remark on the wave function representation of the above basic states.
We have already seen that the two-particle state $\had(\bspsi)\had(\bsph)\vac$ corresponds to the wave function $\{\psi(\bsr_1)\ph(\bsr_2)+\zeta\,\psi(\bsr_2)\ph(\bsr_1)\}/\sqrt{2}$.
Similarly one can show that the wave function representation of the state $\had(\bsph_1)\cdots\had(\bsph_N)\vac$ is
\eq
\Phi(\roN)=\frac{1}{\sqrt{N!}}\sum_P\zeta^P\ph_1(\bsr_{P(1)})\,\ph_2(\bsr_{P(2)})\cdots\ph_N(\bsr_{P(N)}),
\lb{PWF}
\en
where $P$ is summed over all permutations of $\{1,\ldots,N\}$.

\sproof{Proof}
We give inductive proof.\footnote{%
The proof is not interesting.  It may be skipped.
}
The statement is already shown for $N=1$ and 2.
We assume that it is valid for $N-1$, i.e., the wave function representation of the $(N-1)$ particle state $\had(\bsph_2)\cdots\had(\bsph_N)\vac$ is
\eq
\Phi'(\bsr_2,\ldots,\bsr_N)=\frac{1}{\sqrt{(N-1)!}}\sum_{P'}\zeta^{P'}\ph_2(\bsr_{P'(2)})\,\ph_3(\bsr_{P'(3)})\cdots\ph_N(\bsr_{P'(N)}),
\en
where $P'$ is summed over permutations of $\{2,3,\ldots,N\}$.
By using the basic definition \rlb{CA3}, the desired wave function representation of $\had(\bsph_1)\had(\bsph_2)\cdots\had(\bsph_N)\vac$ is written as
\eqa
\Phi(\bsr_1,\ldots,\bsr_N)&=\frac{1}{\sqrt{N}}\sum_{j=1}^N
\zeta^{j-1}\,\ph_1(\bsr_j)\,\Phi'(\bsr_1,\ldots,\breve{\bsr}_j,\ldots,\bsr_N)
\nl&=\frac{1}{\sqrt{N!}}
\sum_{j=1}^N\sum_{P'}\zeta^{j-1}\zeta^{P'}\ph_1(\bsr_j)\,\ph_2(\bsr_{P'_j(2)})\,\ph_3(\bsr_{P'_j(3)})\cdots\ph_N(\bsr_{P'_j(N)}).
\ena
Here, in order to incorporate the relabelling $(\bsr_2,\ldots,\bsr_N)\to(\bsr_1,\ldots,\breve{\bsr}_j,\ldots,\bsr_N)$ of the variables, we defined
\eq
P'_j(k)=\begin{cases}
P'(k)-1&\text{if $P'(k)\le j$},\\
P'(k)&\text{if $P'(k)>j$}.
\end{cases}
\en
We can now define permutation $P$ of $\{1,\ldots,N\}$ by $P(1)=j$ and $P(k)=P'_j(k)$ for $k=2,\ldots,N$.
Since one finds $\zeta^{j-1}\zeta^{P'}=\zeta^P$ by inspection, we get the desired expression \rlb{PWF}.~\qedm

\paragraph*{Slater determinant states for fermions}
For fermions, where we set $\zeta=-1$, the state $\had(\bsph_1)\cdots\had(\bsph_N)\vac$ is called the Slater determinant state.
This is because its wave function representation \rlb{PWF} has a precise form of determinant.
Slater determinant states, although being very special $N$ particle states, play important roles in the theory of many fermions.
Let us state two important properties of Slater determinant states.

The first property is about the essential role of linear independence of the single-particle states $\bsph_1,\ldots,\bsph_N$.
\begin{T}\label{t:Slater1}
A Slater determinant state $\ket{\Phi}=\had(\bsph_1)\cdots\had(\bsph_N)\vac$ is nonzero if and only if the states $\bsph_1,\ldots,\bsph_N$ are linearly independent.
\end{T}

\nproof{Proof}
We note that \rlb{PhiPsi} implies $\braket{\Phi}{\Phi}=\det[\sfG]$, where $\sfG$ is the Gramm matrix associated with $\bsph_1,\ldots,\bsph_N$, i.e., an $N\times N$ matrix defined by $(\sfG)_{j,k}=\sbkt{\bsph_j,\bsph_k}$.
It is a well-known theorem in linear algebra that $\det[\sfG]$ is nonzero if and only if $\bsph_1,\ldots,\bsph_N$ are linearly independent.~\qedm

\bigskip
The second property shows that a Slater determinant state $\had(\bsph_1)\cdots\had(\bsph_N)\vac$ is fully determined by the subspace spanned by the states $\bsph_1,\ldots,\bsph_N$.
\begin{T}
Suppose that the two sets $\{\bsph_1,\ldots,\bsph_N\}$ and $\{\bspsi_1,\ldots,\bspsi_N\}$ of states in $\calH_1$ span the same $N$-dimensional subspace of $\calH_1$.
Then there is a nonzero constant $c\in\bbC$, and we have
\eq
\had(\bsph_1)\cdots\had(\bsph_N)\vac=c\,\had(\bspsi_1)\cdots\had(\bspsi_N)\vac,
\en
i.e., the two Slater determinant states are the same.
\end{T}

\nproof{Proof}
From the assumption we have $\bsph_j=\sum_{k=1}^N\beta_{j,k}\,\bspsi_k$ with some $\beta_{j,k}\in\bbC$ for any $j=1,\ldots,N$.
Then we have
\eq
\had(\bsph_1)\cdots\had(\bsph_N)=\sum_{k_1,\ldots,k_N=1}^N\beta_{1,k_1}\cdots\beta_{N,k_N}\,
\had(\bspsi_{k_1})\cdots\had(\bspsi_{k_N}).
\en
Since $\{\had(\bspsi_k)\}^2=0$, the product $\had(\bspsi_{k_1})\cdots\had(\bspsi_{k_N})$ is either zero or $\pm \had(\bspsi_1)\cdots\had(\bspsi_N)$.
We have thus proved that $\had(\bsph_1)\cdots\had(\bsph_N)=c\,\had(\bspsi_1)\cdots\had(\bspsi_N)$ with some $c\in\bbC$.
But Theorem~\ref{t:Slater1} guarantees that $c\ne0$.~\qedm

\paragraph*{Complete orthonormal systems}
%We shall  discuss complete orthonormal systems of $\calH_N$ by using the creation operators.
We fix an arbitrary complete orthonormal system $\{\bsxi_\alpha\}_{\alpha=1,2,\ldots}$ of $\hilb$, and again set $\had_{\alpha}\coloneqq{}\had(\bsxi_\alpha)$.
Then any state in $\calH_N$ is written as a superposition of the states\footnote{%
In Feynman's notation \cite{Feynman} this is $\ket{\alpha_1,\alpha_2,\ldots,\alpha_N}$.
}
\eq
\ket{\Xi_{\alpha_1,\alpha_2,\ldots,\alpha_N}}\coloneqq{}\had_{\alpha_1}\had_{\alpha_2}\cdots\had_{\alpha_N}\vac,
\lb{Xi}
\en
with $\alpha_1,\alpha_2,\ldots,\alpha_N=1,2,\ldots$.
To avoid overcounting, we only consider $\alpha_1,\alpha_2,\ldots,\alpha_N$ such that $\alpha_1\le\alpha_2\le\cdots\le\alpha_N$ for bosons, and $\alpha_1<\alpha_2<\cdots<\alpha_N$ for fermions.
We find from \rlb{PhiPsi} that $\braket{\Xi_{\alpha_1,\ldots,\alpha_N}}{\Xi_{\alpha'_1,\ldots,\alpha'_N}}=0$ unless $(\alpha_1,\ldots,\alpha_N)=(\alpha'_1,\ldots,\alpha'_N)$, and also that
\eq
\braket{\Xi_{\alpha_1,\ldots,\alpha_N}}{\Xi_{\alpha_1,\ldots,\alpha_N}}=
\begin{cases}
\displaystyle\prod_\alpha n_\alpha!&\text{for bosons},\\
1&\text{for fermions}.
\end{cases}
\en
Here $n_\alpha$ is the number of $j$ such that $\alpha_j=\alpha$, which clearly satisfies $\sum_{\alpha=1}^\infty n_\alpha=N$.
We here use the convention 0!=1.

Thus a complete orthonormal system of $\calH_N$ is formed by the states $(\prod_\alpha n_\alpha!)^{-1/2}\ket{\Xi_{\alpha_1,\ldots,\alpha_N}}$ for any $\alpha_1,\ldots,\alpha_N=1,2,\ldots$ such that  $\alpha_1\le\alpha_2\le\cdots\le\alpha_N$ for bosons, and by the states $\ket{\Xi_{\alpha_1,\ldots,\alpha_N}}$ for any $\alpha_1,\ldots,\alpha_N=1,2,\ldots$ such that   $\alpha_1<\alpha_2<\cdots<\alpha_N$ for fermions.
Interestingly the completeness condition of these systems can be compactly written as
\eq
\frac{1}{N!}\sum_{\alpha_1,\ldots,\alpha_N=1}^\infty\ket{\Xi_{\alpha_1,\ldots,\alpha_N}}\bra{\Xi_{\alpha_1,\ldots,\alpha_N}}=\hat{1},
\en
where the right-hand side denotes the identity operator on $\calH_N$.
Here we are intentionally overcounting the states by summing over all $\alpha_1,\ldots,\alpha_N=1,2,\ldots$, without restricting their ordering.

Both for fermions and bosons, the state \rlb{Xi} can be rewritten as
\eq
\ket{\Xi_{\alpha_1,\alpha_2,\ldots,\alpha_N}}=(\had_1)^{n_1}(\had_2)^{n_2}(\had_3)^{n_3}\cdots\vac
=\Bigl(\prod_{\alpha=1}^\infty(\had_\alpha)^{n_\alpha}\Bigr)\vac,
\en
where $n_\alpha$ is defined as above.
We have $n_\alpha=0,1,2\ldots$ for bosons and $n_\alpha=0,1$ for fermions.
Note that only a finite number of $n_\alpha$ can be nonzero because of the constraint $\sum_{\alpha=1}^\infty n_\alpha=N$.
This motivates us to define another representation of the basis state\footnote{In Feynman's notation \cite{Feynman} this is $\ket{n_1,n_2,n_3,\ldots}$.}
\eq
\ket{\Gamma_{n_1,n_2,n_3,\ldots}}\coloneqq{}\biggl(\prod_{\alpha=1}^\infty\frac{(\had_\alpha)^{n_\alpha}}{\sqrt{n_\alpha!}}\biggr)\vac,
\lb{Gamma}
\en
where the sequence $n_1,n_2,n_3,\ldots$ takes any combinations of allowed values with the constraint $\sum_{\alpha=1}^\infty n_\alpha=N$.
These states clearly form a complete orthonormal system of $\calH_N$.
Note that $n_\alpha$ can be interpreted as the number of particles occupying the single-particle state $\bsxi_\alpha$.
The description of many-particle states in terms of the basis \rlb{Gamma} is known as the occupation number representation.

\paragraph*{The ``second quantization'' of operators}
We still fix an arbitrary complete orthonormal system $\{\bsxi_\alpha\}_{\alpha=1,2,\ldots}$ of $\hilb$, and write $\had_{\alpha}\coloneqq{}\had(\bsxi_\alpha)$.
Let $\ho$ be an arbitrary operator on $\calH_1$.
We then define an operator $\hcB(\ho)$ on $\calH_N$ with any $N$ by 
\eq
\hcB(\ho)=\sum_{\alpha,\beta=1}^\infty\had_\alpha\,\sbkt{\bsxi_\alpha,\ho\,\bsxi_\beta}\,\ha_\beta.
\lb{hcB}
\en
The operator $\hcB(\ho)$ is often called (again misleadingly) the ``second quantization'' of $\ho$, although there is nothing ``quantized''.
It is crucial to note that the operator $\hcB(\ho)$ does not depend on the choice of a complete orthonormal system.
To see this take another arbitrary complete orthonormal system $\{\bskappa_\mu\}_{\mu=1,2,\ldots}$ of $\calH_1$.
Since $\bsxi_\alpha=\sum_{\mu=1}^\infty\bskappa_\mu\,\sbkt{\bskappa_\mu,\bsxi_\alpha}$, we see from the linearity  \rlb{CA4B} of the creation operator  that $\had(\bsxi_\alpha)=\sum_{\mu=1}^\infty\had(\bskappa_\mu)\,\sbkt{\bskappa_\mu,\bsxi_\alpha}$.
Thus we have
\eqa
\hcB(\ho)&=\sum_{\alpha,\beta=1}^\infty\had(\bsxi_\alpha)\,\sbkt{\bsxi_\alpha,\ho\,\bsxi_\beta}\,\ha(\bsxi_\beta)
\nl&=\sum_{\alpha,\beta,\mu,\nu=1}^\infty\had(\bskappa_\mu)\,\sbkt{\bskappa_\mu,\bsxi_\alpha}
\,\sbkt{\bsxi_\alpha,\ho\,\bsxi_\beta}\,\sbkt{\bsxi_\beta,\bskappa_\nu}\,\ha(\bskappa_\nu)
\nl&=\sum_{\mu,\nu=1}^\infty\had(\bskappa_\mu)\,\sbkt{\bskappa_\mu,\ho\,\bskappa_\nu}\,\ha(\bskappa_\nu),
\ena
which shows the desired independence.

For an arbitrary $\bsph\in\calH_1$, we find from the (anti)commutation relations \rlb{comm} that
\eqa
\hcB(\ho)\,\had(\bsph)&=\sum_{\alpha,\beta=1}^\infty\had_\alpha\,\sbkt{\bsxi_\alpha,\ho\,\bsxi_\beta}\,\ha_\beta\,\had(\bsph)
\nl&=\zeta\sum_{\alpha,\beta=1}^\infty\had_\alpha\,\had(\bsph)\,\sbkt{\bsxi_\alpha,\ho\,\bsxi_\beta}\,\ha_\beta
+\sum_{\alpha,\beta=1}^\infty\had_\alpha\,\sbkt{\bsxi_\alpha,\ho\,\bsxi_\beta}\,\sbkt{\bsxi_\beta,\bsph}
\nl&=\had(\bsph)\sum_{\alpha,\beta=1}^\infty\had_\alpha\,\sbkt{\bsxi_\alpha,\ho\,\bsxi_\beta}\,\ha_\beta
+\sum_{\alpha=1}^\infty\had_\alpha\,\sbkt{\bsxi_\alpha,\ho\,\bsph}.
\ena
From the linearity \rlb{CA4B}, we see that the second term in the right-hand side is
\eq
\sum_{\alpha=1}^\infty\had(\bsxi_\alpha)\,\sbkt{\bsxi_\alpha,\ho\,\bsph}
=\had\Bigl(\sum_{\alpha=1}^\infty\bsxi_\alpha\,\sbkt{\bsxi_\alpha,\ho\,\bsph}\Bigr)
=\had(\ho\bsph).
\en
We have thus found an interesting and useful commutation relation
\eq
[\hcB(\ho),\had(\bsph)]=\had(\ho\bsph),
\lb{Bad}
\en
both for bosons and fermions.
It is also a good exercise to show the relation
\eq
[\hcB(\ho),\hcB(\ho')]=\hcB([\ho,\ho']),
\en
for any operators $\ho$ and $\ho'$ on $\calH_1$.

By using \rlb{Bad} repeatedly and noting $\hcB(\ho)\vac=0$, we see for any $\bsph_1,\ldots,\bsph_N\in\calH_1$ that
\eq
\hcB(\ho)\,\had(\bsph_1)\cdots\had(\bsph_N)\,\vac
=\sum_{j=1}^N\had(\bsph_1)\cdots\had(\bsph_{j-1})\,\had(\ho\bsph_j)\,\had(\bsph_{j+1})\cdots\had(\bsph_N)\,\vac.
\lb{Boppp}
\en
Note that the operator $\ho$ is applied to each of the $N$ particles on the right-hand side.
Thus, when acting on $\calH_N$, the ``second quantization'' $\hcB(\ho)$ is interpreted as
\eq
\hcB(\ho)=\ho_1+\ho_2+\cdots+\ho_N,
\lb{o1o2}
\en
where $\ho_j$ is the copy of $\ho$ acting on the $j$-th particle.

A simple but important example is obtained by setting $\ho=\hat{1}$, the identity operator on $\calH_1$.
Then, \rlb{Boppp} reads $\hcB(\hat{1})\,\had(\bsph_1)\cdots\had(\bsph_N)\,\vac=N\,\had(\bsph_1)\cdots\had(\bsph_N)\,\vac$, which means $\hcB(\hat{1})\,\ket{\Phi}=N\,\ket{\Phi}$ for any $\ket{\Phi}\in\calH_N$.
We therefore write $\hN\coloneqq{}\hcB(\hat{1})$,
%\eq
%\hN\coloneqq{}\hcB(\hat{1}),
%\en
and call it the number operator.
From the definition \rlb{hcB}, we have
\eq
\hN=\sum_{\alpha=1}^\infty\had(\bsxi_\alpha)\,\ha(\bsxi_\alpha)=\sum_{\alpha=1}^\infty\hn(\bsxi_\alpha),
\en
where we defined
\eq
\hn(\bsph)=\had(\bsph)\ha(\bsph),
\en
for any $\bsph\in\calH_1$ with $\snorm{\bsph}=1$.
It is easily found that the eigenvalues of $\hn(\bsph)$ are $0,1,2,\ldots$ for bosons, and $0,1$ for fermions.\footnote{%
For bosons, we only need to note that \rlb{comm} implies $[\ha(\bsph),\had(\bsph)]=1$, and repeat the standard argument for the quantum harmonic oscillator.
For fermions, we first notes that $\{\hn(\bsph)\}^2=\had(\bsph)\ha(\bsph)\had(\bsph)\ha(\bsph)=\had(\bsph)\{1-\had(\bsph)\ha(\bsph)\}\ha(\bsph)=\had(\bsph)\ha(\bsph)=\hn(\bsph)$.
Since any eigenvalue $n$ should satisfy the same relation $n^2=n$, we see that $n$ must be 0 or 1.
}
Thus $\hn(\bsph)$ is interpreted as the operator that counts the number of particles in the single-particle state $\bsph$.
By using \rlb{pdad}, one can also check that
\eq
\hN=\int d^3\bsx\,\hpsid(\bsx)\,\hpsi(\bsx)=\int d^3\bsx\,\hat{\rho}(\bsx),
\en
where $\hat{\rho}(\bsx)=\hpsid(\bsx)\,\hpsi(\bsx)$ is called the density operator at $\bsx\in\Rt$.

\paragraph*{Fock space}
The creation and annihilation operators $\had(\bsph)$, $\ha(\bsph)$ and the ``second quantized'' operator $\hcB(\ho)$ act on the $N$ particle Hilbert space $\calH_N$ with any $N=0,1,\ldots$.
Then one may also regard them as operators acting on the Hilbert space
\eq
\calF=\calH_0\oplus\calH_1\oplus\calH_2\oplus\calH_3\oplus\cdots,
\en
which contains states with various particle numbers.
The space $\calF$ is known as the Fock space.
The Fock space $\calF$ consists of states of the form $c_0+\ket{\Phi^{(1)}}+\ket{\Phi^{(2)}}+\ket{\Phi^{(3)}}+\cdots$, where $c_0\in\bbC\cong\calH_0$ is an arbitrary complex number, and $\ket{\Phi^{(N)}}$ is an arbitrary state (which may be vanishing) in $\calH_N$.\footnote{%
To be rigorous the space consists of all possible infinite sums such that $|c_0|^2+\sum_{n=1}^\infty\snorm{\Phi^{(n)}}^2<\infty$.
The reader familiar with the standard formulation of direct sum may better interpret $\ket{\Phi^{(N)}}$ in the above sum as $(0,\ldots,0,\ket{\Phi^{(N)}},0,\ldots)$.
}
For any $N\ne M$, we define $\braket{\Phi^{(N)}}{\Phi^{(M)}}=0$ for any $\ket{\Phi^{(N)}}\in\calH_N$ and $\ket{\Phi^{(M)}}\in\calH_M$.
Equivalently, we can also say that the Fock space $\calF$ consists of all possible linear combinations of states of the form $\had(\bsph_1)\cdots\had(\bsph_N)\vac$ with any $N=0,1,2,\ldots$ and any $\bsph_1,\ldots,\bsph_N\in\calH_1$.

Note that, in a system of electrons or atoms, it is unphysical to consider superpositions of states with different particle numbers.
Such states can never be realized or, more precisely, can never be observed.
The Fock space therefore should be regarded as a purely theoretical object.
Nevertheless, states in the Fock space can be useful in some problems, notably mean-field theories for superconductivity and Bose-Einstein condensation.

\section{Schr\"odinger equation and Hamiltonians}
The Schr\"odinger equation (in the wave function representation) of interacting (non-relativistic) particles is
\eq
(\hH_0+\hH_{\rm int})\,\Phi(\roN)=E\,\Phi(\roN).
\en
The Hamiltonian of a non-interacting system is
\eq
\hH_0=\sum_{j=1}^N\Bigl\{\frac{(\hat{\bsp}_j)^2}{2m}+V(\hat{\bsr}_j)\Bigr\},
\lb{H0}
\en
where $V(\bsr)$ is the single-particle potential.
The interaction Hamiltonian is
\eq
\hH_{\rm int}={\mathop{\sum_{j,j'=1}^N}_{(j<j')}}
V_{\rm int}(\hat{\bsr}_j-\hat{\bsr}_{j'}),
\lb{Hint}
\en
where $V_{\rm int}(\bsr-\bsr')$ denotes the two-body interaction potential.
Let us rewrite these Hamiltonians using the creation and annihilation operators.

\paragraph*{Non-interacting Hamiltonian}
Comparing \rlb{H0} with \rlb{o1o2}, one immediately finds that $\hH_0=\hcB(\hh)$, where 
\eq
\hh=\frac{\hat{\bsp}^2}{2m}+V(\hat{\bsr})
\en
is the single-particle Hamiltonian.
We shall derive the standard expression of $\hH_0$ in terms of $\hpsi(\bsx)$.
Note first that
\eq
\sbkt{\bsxi_\alpha,\hh\,\bsxi_\beta}=\int d^3\bsx\,\{\xi_\alpha(\bsx)\}^*\Bigl\{-\frac{\hbar^2}{2m}\Lap+V(\bsx)\Bigr\}\xi_\beta(\bsx).
\en
Then, from \rlb{hcB}, we find
\eqa
\hH_0
&=\sum_{\alpha,\beta=1}^\infty\had(\bsxi_\alpha)\,\sbkt{\bsxi_\alpha,\hh\,\bsxi_\beta}\,\ha(\bsxi_\beta)
\nl&=\sum_{\alpha,\beta=1}^\infty\int d^3\bsx\,\had(\bsxi_\alpha)\,\{\xi_\alpha(\bsx)\}^*\Bigl\{-\frac{\hbar^2}{2m}\Lap+V(\bsx)\Bigr\}\xi_\beta(\bsx)\,\ha(\bsxi_\beta)
\nl&=\int d^3\bsx\,\hpsid(\bsx)\Bigl\{-\frac{\hbar^2}{2m}\Lap+V(\bsx)\Bigr\}\hpsi(\bsx)
\nl&=-\frac{\hbar^2}{2m}\int d^3\bsx\,\hpsid(\bsx)\,\Lap\hpsi(\bsx)+\int d^3\bsx\,V(\bsx)\,\hat{\rho}(\bsx),
\lb{H02}
\ena
where we used \rlb{pdad} to get the third line.

Let us also derive the expression in terms of $\hak(\bsk)$.
Substituting \rlb{psdad} into the third expression of  \rlb{H02}, we get
\eqa
\hH_0=&\int d^3\bsx\,\int\frac{d^3\bsk\,d^3\bsk'}{(2\pi)^3}\hakd(\bsk)\,e^{-i\bsk\cdot\bsx}
\Bigl\{-\frac{\hbar^2}{2m}\Lap+V(\bsx)\Bigr\}e^{i\bsk'\cdot\bsx}\,\hak(\bsk')
\nl=&\int d^3\bsx\,\int\frac{d^3\bsk\,d^3\bsk'}{(2\pi)^3}\,e^{-i(\bsk-\bsk')\cdot\bsx}\,\frac{\hbar^2|\bsk'|^2}{2m}\,\hakd(\bsk)\hak(\bsk')
\nl&+\int d^3\bsx\,\int\frac{d^3\bsk\,d^3\bsk'}{(2\pi)^3}\,e^{-i(\bsk-\bsk')\cdot\bsx}\,V(\bsx)\,\hakd(\bsk)\hak(\bsk')
\nl=&\int d^3\bsk\,\frac{\hbar^2|\bsk|^2}{2m}\,\hakd(\bsk)\hak(\bsk)
+\int\frac{d^3\bsk\,d^3\bsk'}{(2\pi)^3}\,\tilde{V}(\bsk-\bsk')\,\hakd(\bsk)\hak(\bsk'),
\lb{H03}
\ena
where we used the expression \rlb{delta} of the delta function, and defined the Fourier transformation of the potential by
\eq
\tilde{V}(\bsk)\coloneqq{}\int d^3\bsx\,e^{-i\bsk\cdot\bsx}\,V(\bsx).
\en
It is illuminating to rewrite the second term as
\eq
\hH_0=\int d^3\bsk\,\frac{\hbar^2|\bsk|^2}{2m}\,\hakd(\bsk)\hak(\bsk)
+\int\frac{d^3\bsk\,d^3\bsq}{(2\pi)^3}\,\tilde{V}(\bsq)\,\hakd(\bsk+\bsq)\hak(\bsk).
\en
This expression shows that a particle with wave number $\bsk$ is annihilated and that with $\bsk+\bsq$ is created, with the amplitude $\tilde{V}(\bsq)$.

Let us assume that the complete orthonormal system $\{\bsxi_\alpha\}_{\alpha=1,2,\ldots}$ consists of the eigenstates of the single-particle Hamiltonian $\hh$, i.e.,
\eq
\hh\,\bsxi_\alpha=\epsilon_\alpha\,\bsxi_\alpha
\lb{hxa}
\en
for any $\alpha=1,2,\ldots$, where $\epsilon_\alpha$ is the single-particle energy eigenvalue.
Then it follows from \rlb{Bad} that
\eq
[\hH_0,\had_\alpha]=\epsilon_\alpha\,\had_\alpha
\en
and hence
\eq
[\hH_0,\had_{\alpha_1}\cdots\had_{\alpha_N}]=\Bigl(\sum_{j=1}^N\epsilon_{\alpha_j}\Bigr)\,\had_{\alpha_1}\cdots\had_{\alpha_N}.
\en
This relation, with $\hH_0\vac=0$, implies
\eq
\hH_0\,\ket{\Xi_{\alpha_1,\ldots,\alpha_N}}=\Bigl(\sum_{j=1}^N\epsilon_{\alpha_j}\Bigr)\,\ket{\Xi_{\alpha_1,\ldots,\alpha_N}},
\en
where the state $\ket{\Xi_{\alpha_1,\ldots,\alpha_N}}$ is defined in \rlb{Xi}.
We have thus found that the state $\ket{\Xi_{\alpha_1,\ldots,\alpha_N}}$ is the energy eigenstate when there is no interaction between particles.
The same conclusion follows from the expression
\eq
\hH_0=\sum_{\alpha=1}^\infty\epsilon_\alpha\,\had_\alpha\ha_\alpha=\sum_{\alpha=1}^\infty\epsilon_\alpha\,\hn_\alpha,
\lb{H0diag}
\en
which follows from \rlb{hxa} and the definition \rlb{hcB}.
Here $\hn_\alpha=\hn(\bsxi_\alpha)$ is the number operator for the state $\bsxi_\alpha$.
The expression \rlb{H0diag} is the well-known diagonalized form of $H_0$.

Assume, for simplicity, that the energy eigenvalues of $\hh$ are non-degenerate and ordered as $\epsilon_\alpha<\epsilon_{\alpha+1}$ for $\alpha=1,2,\ldots$.
Then the ground state of the non-interacting Hamiltonian $\hH_0$ is obtained by minimizing the energy eigenvalue $\sum_{j=1}^N\epsilon_{\alpha_j}$.
For bosons, this is realized when $\alpha_j=1$ for $j=1,\ldots,N$, and hence the ground state and the ground state energy are $\ket{\Xi_{1,1,\ldots,1}}$ and  $N\,\epsilon_1$, respectively.
For fermions, the minimum is realized if we take $\alpha_j=j$ for $j=1,\ldots,N$, and the ground state and the ground state energy are $\ket{\Xi_{1,2,\ldots,N}}$ and $\sum_{\alpha=1}^N\epsilon_{\alpha}$, respectively.

\paragraph*{Time evolution of the annihilation operator in non-interacting systems}
Let $t$ be the time variable, and consider the annihilation operator in the Heisenberg representation given by
\eq
\hpsi(\bsx,t)\coloneqq{}e^{i\hH_0t}\,\hpsi(\bsx)\,e^{-i\hH_0t}.
\lb{psixt}
\en
Recalling that $\frac{d}{dt}e^{i\hH_0t}=i\,e^{i\hH_0t}\hH_0$ and $\frac{d}{dt}e^{-i\hH_0t}=-i\,\hH_0\,e^{i\hH_0t}$, we find that the time derivative of \rlb{psixt} can be written as
\eq
i\frac{\partial}{\partial t}\hpsi(\bsx,t)
=e^{i\hH_0t}\,[\hpsi(\bsx),\hH_0]\,e^{-i\hH_0t}.
\lb{psixt2}
\en
By using the (anti)commutation relations \rlb{pp} and the expression in the third line of \rlb{H02}, one finds that
\eq
[\hpsi(\bsx),\hH_0]=-\frac{\hbar^2}{2m}\Lap\hpsi(\bsx)+V(\bsx)\,\hpsi(\bsx).
\en
Note that this commutation relation is valid for both bosonic and fermionic systems.
Substituting this into \rlb{psixt2}, and using \rlb{psixt}, we observe that
\eq
i\frac{\partial}{\partial t}\hpsi(\bsx,t)
=-\frac{\hbar^2}{2m}\Lap\hpsi(\bsx,t)+V(\bsx)\,\hpsi(\bsx,t).
\lb{psixt3}
\en
Needless to say, the equation has exactly the same form as the Schr\"odinger equation for a single particle.
This similarity, which holds only in non-interacting systems, may be a reason which caused (and still causes) the misunderstanding that $\hpsi(\bsx,t)$ is obtained by ``second quantizing'' the wave function $\psi(\bsx,t)$.

\paragraph*{Interaction Hamiltonian}
The interaction Hamiltonian \rlb{Hint} can be written in terms of the creation and annihilation operators as
\eq
\hH_{\rm int}=\frac{1}{2}\int d^3\bsx\,d^3\bsy\,\hpsid(\bsx)\,\hpsid(\bsy)\,V_{\rm int}(\bsx-\bsy)\,\hpsi(\bsy)\,\hpsi(\bsx).
\lb{Hint2}
\en
To see that this expression correctly recovers the desired \rlb{Hint}, we shall examine its action on the state $\hpsid(\bsx_1)\cdots\hpsid(\bsx_N)\vac$, in which $N$ particles have definite positions $\bsx_1,\ldots,\bsx_N$.
Note first that 
\eq
\hpsi(\bsx)\,\hpsid(\bsx_1)\cdots\hpsid(\bsx_N)\vac
=\sum_{j=1}^N\zeta^{j-1}\,\delta(\bsx-\bsx_j)\,\underbrace{\hpsid(\bsx_1)\cdots\hpsid(\bsx_N)}_{\text{no $\hpsid(\bsx_j)$}}\vac
\en
where we used the (anti)commutation relations \rlb{pp}.
Similarly we find
\eq
\hpsi(\bsy)\,\hpsi(\bsx)\,\hpsid(\bsx_1)\cdots\hpsid(\bsx_N)\vac
={\mathop{\sum_{j,j'=1}^N}_{(j\ne j')}}\eta_{j,j'}\,\delta(\bsx-\bsx_j)\,\delta(\bsy-\bsx_{j'})\,\underbrace{\hpsid(\bsx_1)\cdots\hpsid(\bsx_N)}_{\text{no $\hpsid(\bsx_j), \hpsid(\bsx_{j'})$}}\vac,
\en
where $\eta_{j,j'}=\zeta^{j-1}\zeta^{j'-1}$ if $j>j'$ and $\eta_{j,j'}=\zeta^{j-1}\zeta^{j'}$ if $j<j'$.
Noting that $\hpsid(\bsx)\,\delta(\bsx-\bsx_j)=\hpsid(\bsx_j)\,\delta(\bsx-\bsx_j)$, we have
\eqa
\hpsid(\bsx)\,&\hpsid(\bsy)\,\hpsi(\bsy)\,\hpsi(\bsx)\,\hpsid(\bsx_1)\cdots\hpsid(\bsx_N)\vac
\nl&={\mathop{\sum_{j,j'=1}^N}_{(j\ne j')}}\eta_{j,j'}\,\delta(\bsx-\bsx_j)\,\delta(\bsy-\bsx_{j'})\,
\hpsid(\bsx_j)\,\hpsid(\bsx_{j'})\,\underbrace{\hpsid(\bsx_1)\cdots\hpsid(\bsx_N)}_{\text{no $\hpsid(\bsx_j), \hpsid(\bsx_{j'})$}}\vac
\nl&={\mathop{\sum_{j,j'=1}^N}_{(j\ne j')}}\delta(\bsx-\bsx_j)\,\delta(\bsy-\bsx_{j'})\,
\hpsid(\bsx_1)\cdots\hpsid(\bsx_N)\vac.
\ena
Thus, by applying $\hH_{\rm int}$ in the form of \rlb{Hint2} onto the state $\hpsid(\bsx_1)\cdots\hpsid(\bsx_N)\vac$, and performing the integration, we get
\eq
\hH_{\rm int}\,\hpsid(\bsx_1)\cdots\hpsid(\bsx_N)\vac=\frac{1}{2}\mathop{\sum_{j,j'=1}^N}_{(j\ne j')}V_{\rm int}(\bsx_j-\bsx_{j'})\,\hpsid(\bsx_1)\cdots\hpsid(\bsx_N)\vac,
\en
which clearly coincides with the action of \rlb{Hint}.

Let us finally rewrite \rlb{Hint2} by using $\hakd(\bsk)$ and $\hak(\bsk)$.
Substituting \rlb{psdad} to \rlb{Hint2}, we get
\eqa
&\hH_{\rm int}=
\frac{1}{2(2\pi)^6}
\int d^3\bsx\,d^3\bsy\Bigl(\prod_{\nu=1}^4d^3\bsk_\nu\Bigr)
\,e^{-i\bsk_1\cdot\bsx-i\bsk_2\cdot\bsy+i\bsk_3\cdot\bsy+i\bsk_4\cdot\bsx}\,V_{\rm int}(\bsx-\bsy)\,
\hakd(\bsk_1)\,\hakd(\bsk_2)\,\hak(\bsk_3)\,\hak(\bsk_4)\notag
\intertext{%
By changing the variable of integration from $\bsx$ to $\bsw=\bsx-\bsy$, this becomes
}
&=\frac{1}{2(2\pi)^6}\int d^3\bsw\,d^3\bsy\Bigl(\prod_{\nu=1}^4d^3\bsk_\nu\Bigr)
\,e^{-i(\bsk_1+\bsk_2-\bsk_3-\bsk_4)\cdot\bsy-i(\bsk_1-\bsk_4)\cdot\bsw}\,V_{\rm int}(\bsw)\,
\hakd(\bsk_1)\,\hakd(\bsk_2)\,\hak(\bsk_3)\,\hak(\bsk_4)\nl
&=\frac{1}{2(2\pi)^3}\int\Bigl(\prod_{\nu=1}^4d^3\bsk_\nu\Bigr)
\delta(\bsk_1+\bsk_2-\bsk_3-\bsk_4)\,\tilde{V}_{\rm int}(\bsk_1-\bsk_4)\,
\hakd(\bsk_1)\,\hakd(\bsk_2)\,\hak(\bsk_3)\,\hak(\bsk_4),
\lb{HInt3}
\ena
where we defined the Fourier transformation of the interaction potential as 
\eq
\tilde{V}_{\rm int}(\bsk)=\int d^3\bsx\,e^{-i\bsk\cdot\bsx}\,V_{\rm int}(\bsx).
\en
Note that the final expression in \rlb{HInt3} shows that the total momenta before and after the action of $V_{\rm int}$ are conserved, i.e.,  $\bsp_1+\bsp_2=\bsp_3+\bsp_4$, where $\bsp_\nu=\hbar\bsk_\nu$.
By performing the integration over $\bsk_1$, writing $\bsk=\bsk_4$, $\bsk'=\bsk_3$, $\bsq=\bsk_3-\bsk_2$, and changing the variable of integration from $\bsk_2$ to $\bsq$, the expression \rlb{HInt3} is rewritten as
\eq
\Hint=\frac{1}{2(2\pi)^3}\int d^3\bsk\,d^3\bsk'\,d^3\bsq\,\tilde{V}_{\rm int}(\bsq)\,
\hakd(\bsk+\bsq)\,\hakd(\bsk'-\bsq)\,\hak(\bsk')\,\hak(\bsk),
\en
which shows that momentum $\hbar\bsq$ is transformed from one particle to the other with amplitude $\tilde{V}_{\rm int}(\bsq)$.

\bigskip\small
I wish to thank Daisuke Ida for his valuable comments and suggestions and the undergraduate students in our theoretical physics group for their careful readings of the manuscript and useful comments.
I also thank Yu Nakayama and Mizuki Yamaguchi for their helpful comments on the discussion in Section~1 about the symmetry of multi-particle wave functions.

\end{document}

%% file: macros.tex
\newlength{\mycm}

\newcommand{\nl}{\notag\\}

\makeatletter
\@addtoreset{equation}{section}
\makeatother

\def\eq#1\en{\begin{equation}#1\end{equation}}  
\def\eqsplit#1\ensplit{
	\begin{equation}\begin{split}#1\end{split}\end{equation}
	}
\def\eqalign#1\enalign{
	\begin{align}#1\end{align}
	}
\def\eqa#1\ena{
	\begin{align}#1\end{align}
	}
\def\eqg#1\eng{
	\begin{gather}#1\end{gather}
}
\def\eqmul#1\enmul{
	\begin{multline}#1\end{multline}
	}
\newcommand{\lb}[1]  {\label{e:#1}}
\newcommand{\rlb}[1] {\eqref{e:#1}}     
\newtheorem{theorem}{Theorem}[section]
\newtheorem{T}[theorem]{Theorem}

\newcommand{\nproof}[1]{\par\noindent{\em #1\/}:}
\newcommand{\sproof}[1]{\par\bigskip\noindent{\em #1\/}:}
\newcommand{\qedm}{\rule{1.5mm}{3mm}}
\newcommand{\abs}[1]{\left|#1\right|}

\newcommand{\snorm}[1]{\Vert#1\Vert}

\newcommand{\sbkt}[1]{\langle#1\rangle}

\newcommand{\bra}[1]{\langle#1|}
\newcommand{\ket}[1]{|#1\rangle}
\newcommand{\braket}[2]{\langle#1|#2\rangle}

\newcommand{\sumtwo}[2]%
{\mathop{\sum_{#1}}_{#2}}
\newcommand{\sumthree}[3]%
{\mathop{\mathop{\sum_{#1}}_{#2}}_{#3}}
\newcommand{\sumfour}[4]%
{\mathop{\mathop{\mathop{\sum_{#1}}_{#2}}_{#3}}_{#4}} 
\newcommand{\prodtwo}[2]%
{\mathop{\prod_{#1}}_{#2}}
\newcommand{\mintwo}[2]%
{\mathop{\min_{#1}}_{#2}}
\newcommand{\maxtwo}[2]%
{\mathop{\max_{#1}}_{#2}}
\newcommand{\maxthree}[3]%
{\mathop{\mathop{\max_{#1}}_{#2}}_{#3}}
\newcommand{\limtwo}[2]%
{\mathop{\lim_{#1}}_{#2}}
\newcommand{\suptwo}[2]%
{\mathop{\sup_{#1}}_{#2}}
\newcommand{\supthree}[3]%
{\mathop{\mathop{\sup_{#1}}_{#2}}_{#3}}
\newcommand{\supfour}[4]%
{\mathop{\mathop{\mathop{\sup_{#1}}_{#2}}_{#3}}_{#4}} 
\newcommand{\inftwo}[2]%
{\mathop{\inf_{#1}}_{#2}}
\newcommand{\infthree}[3]%
{\mathop{\mathop{\inf_{#1}}_{#2}}_{#3}}
\newcommand{\inffour}[4]%
{\mathop{\mathop{\mathop{\inf_{#1}}_{#2}}_{#3}}_{#4}} 
\newcommand\calB{{\cal B}}

\newcommand\calF{{\cal F}}

\newcommand\calH{{\cal H}}

\newcommand{\bsk}{\boldsymbol{k}}

\newcommand{\bsp}{\boldsymbol{p}}
\newcommand{\bsq}{\boldsymbol{q}}
\newcommand{\bsr}{\boldsymbol{r}}

\newcommand{\bsu}{\boldsymbol{u}}

\newcommand{\bsw}{\boldsymbol{w}}
\newcommand{\bsx}{\boldsymbol{x}}
\newcommand{\bsy}{\boldsymbol{y}}
\newcommand{\bsz}{\boldsymbol{z}}
\newcommand{\bseta}{\boldsymbol{\eta}}

\newcommand{\bsph}{\boldsymbol{\varphi}}
\newcommand{\bspsi}{\boldsymbol{\psi}}

\newcommand{\bsxi}{\boldsymbol{\xi}}

\newcommand{\ha}{\hat{a}}
\newcommand{\had}{\hat{a}^\dagger}

\newcommand{\hh}{\hat{h}}
\newcommand{\ho}{\hat{o}}

\newcommand{\hn}{\hat{n}}

\newcommand{\hA}{\hat{A}}
\newcommand{\hB}{\hat{B}}

\newcommand{\hH}{\hat{H}}

\newcommand{\hN}{\hat{N}}

\newcommand{\sfA}{\mathsf{A}}

\newcommand{\sfG}{\mathsf{G}}

\newcommand{\bbC}{\mathbb{C}}

\newcommand{\bbR}{\mathbb{R}}

\newcommand{\Lap}{\boldsymbol{\bigtriangleup}}

\newcommand{\dmat}[1]{\abs{\begin{matrix}#1\end{matrix}}}

{\end{itemize}\end{FRAME}}
\newcounter{mondaiCounter}[subsection]
\renewcommand{\themondaiCounter}{\thesubsection.\alph{mondaiCounter}}
\newcounter{mondaiCounterS}[section]
\renewcommand{\themondaiCounterS}{\thesection.\alph{mondaiCounterS}}
\newcommand{\toi}[1]{
\ifthenelse{\value{subsection}=0}
{\refstepcounter{mondaiCounterS}
\par\bigskip\noindent\underline{Problem \themondaiCounterS}\label{#1}~(\hyperlink{s:#1}{solution$\to$})~}
{\refstepcounter{mondaiCounter}
\par\bigskip\noindent\underline{Problem \themondaiCounter}\label{#1}~(\hyperlink{s:#1}{solution$\to$})~}
}
\newcommand{\toin}[1]{
\ifthenelse{\value{subsection}=0}
{\refstepcounter{mondaiCounterS}
\par\noindent\underline{Problem \themondaiCounterS}\label{#1}~(\hyperlink{s:#1}{solution$\to$})~}
{\refstepcounter{mondaiCounter}
\par\noindent\underline{Problem \themondaiCounter}\label{#1}~(\hyperlink{s:#1}{solution$\to$})~}
}

\newcounter{reiCounter}[subsection]
\renewcommand{\thereiCounter}{\thesubsection.\alph{reiCounter}}
\newcounter{reiCounterS}[section]
\renewcommand{\thereiCounterS}{\thesection.\alph{reiCounterS}}
\newcommand{\rei}[1]{
\ifthenelse{\value{subsection}=0}
{\refstepcounter{reiCounterS}
\par\bigskip\noindent{\bf Example \thereiCounterS}\label{#1}~}
{\refstepcounter{reiCounter}
\par\bigskip\noindent{\bf Exmaple \thereiCounter}\label{#1}~}
}
\newcommand{\vac}{\ket{\Phi_\mathrm{vac}}}
\newcommand{\vacb}{\bra{\Phi_\mathrm{vac}}}

\newcommand{\hilb}{\mathfrak{h}}
\newcommand{\bss}{\boldsymbol{\sigma}}

\newcommand{\ph}{\varphi}

\newcommand{\tCd}{{\sfC}^{\dagger}\!}
\newcommand{\tC}{{\sfC}}
\newcommand{\Hint}{\hH_{\rm int}}
\newcommand{\bsPhi}{\boldsymbol{\Phi}}
\newcommand{\bsPsi}{\boldsymbol{\Psi}}
\newcommand{\bsXi}{\boldsymbol{\Xi}}

\newcommand{\calHwf}{\calH^\mathrm{wf}}